\documentclass[12pt]{article}

\input epsf
\newcount\figno
\def\fig#1#2#3{
\par\begingroup\parindent=0pt\leftskip=1cm\rightskip=1cm\parindent=0pt
\baselineskip=11pt
\global\advance\figno by 1
\epsfxsize=#3
\centerline{\epsfbox{#2}}
\vskip 12pt
{\bf Figure \the\figno:} #1\par
\endgroup\par
}
\def\figlabel#1{\xdef#1{\the\figno
\mbox{ }}}
\def\encadremath#1{\vbox{\hrule\hbox{\vrule\kern8pt\vbox{\kern8pt
\hbox{$\displaystyle #1$}\kern8pt}
\kern8pt\vrule}\hrule}}

\def\href#1#2{#2}
\def\beq{\begin{equation}}
\def\eeq{\end{equation}}

\def\drawbox#1#2{\hrule height#2pt 
        \hbox{\vrule width#2pt height#1pt \kern#1pt 
              \vrule width#2pt}
              \hrule height#2pt}
\def\Fund#1#2{\vcenter{\vbox{\drawbox{#1}{#2}}}}
\def\Asym#1#2{\vcenter{\vbox{\drawbox{#1}{#2}
              \kern-#2pt       % line up boxes
              \drawbox{#1}{#2}}}}
 
\def\fund{\Fund{6.5}{0.4}}
\def\asym{\Asym{6.5}{0.4}}

\begin{document}
\begin{titlepage}

\begin{center}
\today
\hfill IASSNS-HEP-98/2\\
\hfill HUB-EP 98/1\\
\hfill hep-th/9801017
\vspace{2cm}

{\Large\bf Brane Dynamics and Chiral non-Chiral Transitions}

\vspace{2cm}
{\large Ilka Brunner $\!{}^{\dag}$, Amihay Hanany $\!{}^{*}$, 
Andreas Karch $\!{}^{\dag}$ \\
and\,\,\,
Dieter L\"ust ${}^{\dag}$}

\vspace{1cm}

{\it ${}^*$School of Natural Sciences,
Institute for Advanced Study\\
Olden Lane, Princeton, NJ 08540, USA\\[0.3cm]
${}^{\dag}$Institut f\"ur Physik,
Humboldt Universit\"at Berlin\\
Invalidenstr. 110\\
10115 Berlin, Germany}

\end{center}

\vspace{1.5cm}

\begin{abstract}
We study brane realizations of chiral matter
in $N=1$ supersymmetric gauge theories in four dimensions.
A ``cross" configuration which leads to ``flavor doubling" is found
to have a superpotential.
The main example is realized using a special ``fork" configuration.
Many of the results are found by studying a $SU \times SU$ 
product gauge group first.
The chiral theory is then an orientifold projection of the product gauge group.
An interesting observation in the brane picture is that
there are transitions between chiral and non chiral models.
These transitions are closely related to small instanton 
transitions in six dimensions.
\end{abstract}

\end{titlepage}

\section{Introduction}

The past few years have seen tremendous progress in our
understanding of non-perturbative effects in supersymmetric
field theory and string theory. Many insights into gauge theories
have emerged by using D brane techniques. A particular brane
configuration where D branes are suspended between NS branes
was first used in \cite{HW} and later used to study theories
with eight or four supercharges in various dimensions.
Here, the gauge theory under consideration is realized as the
world volume theory of the lowest dimensional D brane in the setup.
Fundamental matter can be included by adding higher dimensional
D branes. This basic setup can be modified by including further
ingredients
such as orientifolds.
Of particular interest are configurations corresponding to
$ N=1$ supersymmetric theories in four dimensions. 
Here, one success was that Seiberg's duality \cite{seiberg}, could be reproduced
from the brane picture \cite{EGK}. However, a brane configuration
giving $N=1$ differs from a setup leading to $N=2$ only by a
rotation angle of the NS branes with respect to each other.
Therefore, a brane configuration leading to $N=1$ locally looks like
an $N=2$ configuration. An important difference between $N=1$
and $N=2$ is chiral symmetry. A priori we only see a diagonal
subgroup of the full chiral symmetry group in the brane picture.
In \cite{brodie} it was suggested that the full chiral symmetry
is restored if the flavour giving D branes are broken in two
pieces by an NS brane whose world volume directions coincide
with world volume directions of the D6 branes. More generally, chiral
multiplets correspond to ``half-branes'' ending on NS branes.
Anomaly cancellation in field theory translates to RR charge
conservation in the string theory context \cite{hz}.
In this paper we will study the localization of chiral matter
in $N=1$, $d=4$
in some more detail. In the first section we look at situations
where D6 branes touch D4 branes at points where also NS$'$ branes
are localized. Here, we observe the presence of additional 
multiplets from the breaking of the D6 and D4 branes, as well as a nonzero
superpotential. In the next section
we use this observation in the context of product gauge groups.
After that we specialize to configurations which are symmetric
with respect to the location of an NS brane. Here it is
possible to introduce orientifolds in various ways. One of our
models is a chiral theory with chiral multiplets in the
fundamental, symmetric and antisymmetric representation.
This model has also been studied recently in a paper by Landsteiner, Lopez
and Lowe \cite{LLL}, which appeared while this work was in progress.
Other realizations of chiral gauge theories from branes can be
obtained by looking at brane configurations in an
orbifold background \cite{lykken,lykken2}.
Finally, we show in the brane picture that there exists a phase
transition from this chiral model to a non-chiral model.
This is a very striking effect, in which the continuation past
infinite coupling of a non-chiral theory turns out to be the chiral
model.
Such a transition was studied recently in a T-dual model in the brane
realization of six dimensional theories \cite{hztwo}.
 From a T-dual point of view,  of the world volume theory on the D6 branes
this corresponds to a small instanton transition.
Thus, the two field theory transitions are a consequence of the same brane 
motion, which puts them on the same universality class. Such chirality
changing trasitions have been previously considered from 
geometric considerations in \cite{kachru,louis}.

Let us set some notation and recall simple facts which we will use throughout
this paper. 
We work in Type IIA superstring theory.
The various branes considered in this paper will be
NS brane which spans directions 012345,
D4 brane along the directions 01236,
D6 brane with 0123789,
NS$'$ with 012389,
D6$'$ with 0123457.
The presence of these branes breaks supersymmetry to 4 supercharges and lead to
a four dimensional $N=1$ supersymmetic gauge theory on the world volume of the
D4 branes. These branes are finite in the $x^6$ direction and reduction on this
interval leads to the four dimensional theory sitting in 0123.
$SO(1,9)$ space time Lorentz symmetry is broken to $SO(1,3)\times SO(2)\times
SO(2)$. The first factor is the Lorentz symmetry of the four dimensional theory.
The two other $SO(2)$ factors act on 45 and 89 directions and are R-symmetries
of the four dimensional theory.
A rotation of the NS branes and the D6 branes along the 45-89 directions does
not break further the supersymmetry and we will use this property to add more
parameters to the theory studied. 

\section{Localization of Chiral Matter in Space}

In this section we will study a particular configuration in which chiral matter
with a non-trivial superpotential is localized in spacetime.
This configuration serves as a building block for the models that are studied
in this paper and for other generalizations.

\vspace{1cm}
\fig{
A configuration of branes in which a single chiral field is localized in
space time. A D6 brane ending on a NS$'$ brane meets a D4 brane which also ends
on the NS$'$ brane.}
{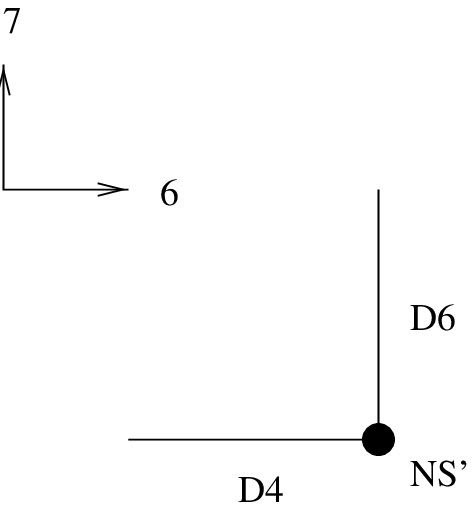}{6truecm}
\figlabel{\ddns}
\vspace{1cm}

Let us start by recalling some of the relevant configurations already studied.
In \cite{brodie}, it was suggested that chiral symmetry is enhanced at special
points were a D6 brane meets a NS$'$ brane. This led to a localization of chiral
matter in the configuration of figure \ddns which arises at the
intersection of a D4 brane and a D6 brane which end on NS$'$ brane \cite{hz}.
Evidence for this proposal using various arguments were given in
\cite{ah,hz,hh}.

\vspace{1cm}
\fig{A ``cross" configuration.
An intersection of a D4 brane, a D6 brane and a NS$'$ brane. There are four
chiral fields, two chiral bi-fundamentals and a superpotential.}
{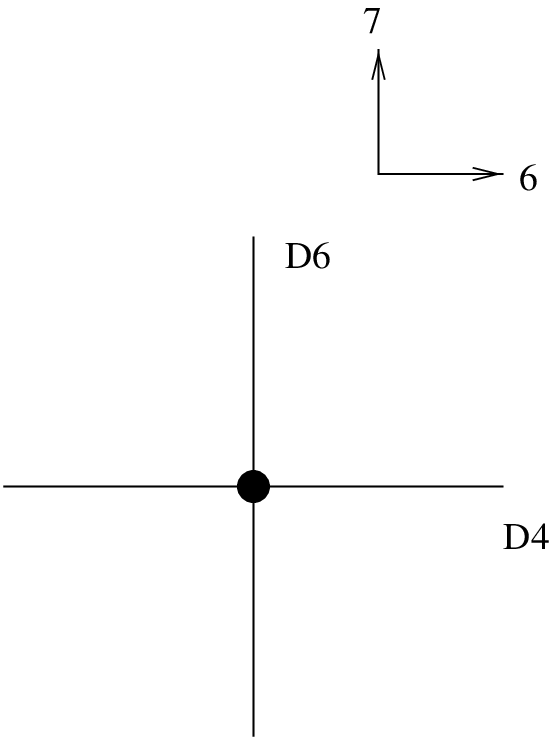}{7truecm}
\figlabel{\cross}
\vspace{1cm}

We wish to use this identification to study the configuration given in figure
\cross. First let us recall that when a D6 brane meets a D4 brane there is
a massless hypermultiplet which transforms under the $U(1)\times U(1)$
gauge groups which sit on the D4 and D6 branes. The number of supersymmetries
for such a configuration is 8 and so there is a superpotentail which is
restricted by the supersymmetry to be $(m-x)\tilde QQ$, where $m$ is the 45
position of the D6 brane, $x$ is the 45 position of the D4 brane and
$\tilde q, q$ transform as $(-1,1)$ and $(1,-1)$, respectively.

We can slowly tune the 
position of a NS$'$ brane to touch the intersection of the
D4 and D6 branes. The number of supersymmetries locally is now 4.
At this point, both the D4 and the D6 branes can break and
the gauge symmetry is enhanced to
$U(1)_u\times U(1)_r\times U(1)_d\times U(1)_l$.
The $u,d$ indices correspond to the two parts of the D6 branes and the $l,r$
indices correspond to the two parts of the D4 branes.
As usual for the transitions involving NS branes which lead to breaking of the
D branes, we should look for an interpretation as a Higgs mechanism. Since,
the number of vector multiplets is increased by two, we need to look for two
more massless chiral fields. By applying the configuration of figure \ddns,
we see that we have four copies of chiral multiplets $\tilde Q, Q, \tilde R, R$.
They carry charges (1,-1,0,0), (0,1,-1,0), (0,0,1,-1), (-1,0,0,1), respectively
under the gauge groups.
In addition there are bi-fundamental fields for the intersection of the two new
D4 branes, $\tilde F, F$ with charges (0,1,0,-1) and (0,-1,0,1), respectively.
Two more bi-fundamental fields come from the two new D6 branes.
For the moment, we will ignore the bi-fundamentals for the D6 branes, since they
have six dimensional kinetic terms and so are not dynamical for the four
dimensional system.
All together we have $4\choose2$ possible fields coming from the intersection of
each pair of branes from possible 4.

\vspace{1cm}
\fig{
A deformation of the ``cross" configuration by reconnecting the two D4 branes
and moving the D4 brane in the 45 direction. The dotted lines do not
represent branes.}
{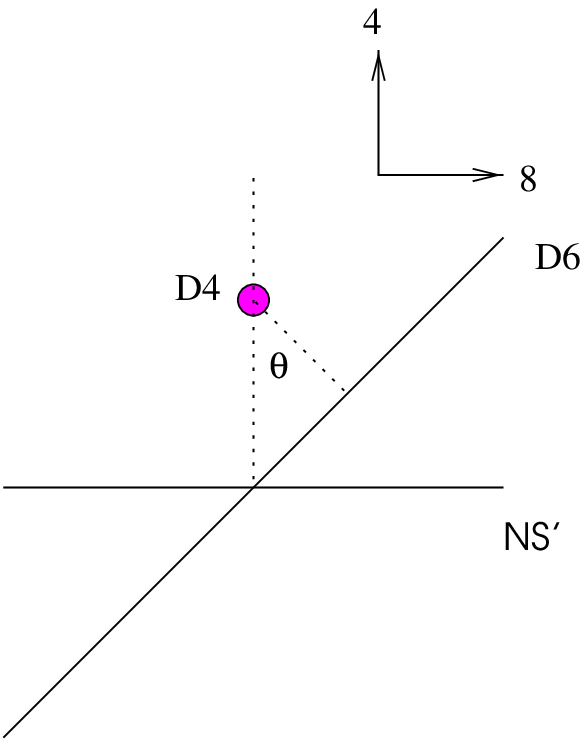}{6truecm}
\figlabel{\defcross}
\vspace{1cm}

The system has 4 supercharges and we can not exclude the possibility of a
superpotential.
We can write down a superpotential of the form
\beq
a \tilde Q\tilde F R + b \tilde R F Q.
\label{supo}
\eeq
The coefficients $a$ and $b$ are determined by looking at various deformations
of this configuration.
One possible deformation is a rotation of the D6 branes in the 45-89 direction.
This rotation does not break further the supersymmetry. There is one angle
between the D6 branes and the NS$'$ brane. One can think that there may be two
angles, one for the upper D6 brane and one for the lower D6 brane. However, RR
charge conservation of the D6 branes implies that they must be parallel to
each other and so the angles are the same. Let us choose an orientation by
fixing the NS$'$ brane and let $\theta$ be the angle between the
rotated D6 brane and the NS$'$ brane. It is defined in such a way that for
$\theta=0$ we recover the D6 brane whereas for $\theta={\pi\over2}$ we get the
D6$'$ brane.

Let us reconnect the two D4 branes and move them in the 45 directions as in
figure \defcross. This motion corresponds to an expectation value for the
bi-fundamental fields $\tilde F$ and $F$. From equation (\ref{supo}) we see that
this gives a mass to all four fields $ \tilde Q, R, \tilde R, Q$. The mass is
given by the distance between the two D branes, $F cos \theta$.
The symmetry of 
the two objects implies that $a$ and $b$ are of the same magnitude
up to a sign. This is determined to be a minus sign by a motion of the D6 branes
along the $x^6$ direction.
The final superpotential is
\beq
\cos \theta \tilde Q\tilde F R - \cos\theta \tilde R F Q.
\label{pupo}
\eeq
In this discussion we have omitted possible chiral fields associated to the
motion of the D4 branes along the NS$'$ branes. Inclusion of these fields
gives rise to interactions which are restricted by supersymmetry with 8
supercharges and contribute appropriate terms to the superpotential.

We can now generalize the construction to configurations with more than one
brane of a given type. Suppose we have $n_l (n_r)$ D4 branes to the left (right)
of the NS$'$ brane and $n_u (n_d)$ D6 branes above (below) the NS$'$ brane.
First, $n_u=n_d=n$ by RR charge conservation.
The gauge and global symmetries are
$U(n)_u\times U(n_r)\times U(n)_d\times U(n_l)$, under which the fields,
$ \tilde Q,Q,  \tilde R,R$ transform as 
$(\fund,\overline{\fund},1,1)$, $(1,\fund,\overline{\fund},1)$,
$(1,1,\fund,\overline{\fund})$, $(\overline{\fund},1,1,\fund)$, together with
bi-fundamental fields
$\tilde F, F$ which transform under $(1,\fund,1,\overline{\fund})$ and
$(1,\overline{\fund},1,\fund)$, respectively.
The superpotential is given by equation (\ref{pupo}).
We can break the global symmetries $U(n)\times U(n)$ by rotating each D6 brane
independently. This breaks the global symmetry generically to $U(1)^{2n}$.
The different angles correspond to the eigenvalues of the Yukawa coupling matrix
in equation (\ref{pupo}).

Equipped with this configuration and its superpotential, let us revisit the
brane configuration for products of $SU$ gauge groups.

\section{Product Gauge Groups}

In this section we will use the ``cross" configuration studied in the previous
section to study a system which describes product of gauge groups. This will
help us understand the doubling of fields at the intersection between the two
gauge groups which was studied in section 2.9 of \cite{ah}.

\vspace{1cm}
\fig{A demonstration of the ``doubling" of matter for a brane configuration of
a product of two gauge groups. The upper and lower NS$'$ branes are put
sufficiently far from the 4d system. They are introduced to study the six
dimensional system which governs the global symmetry of the four dimensional
thoery.}
{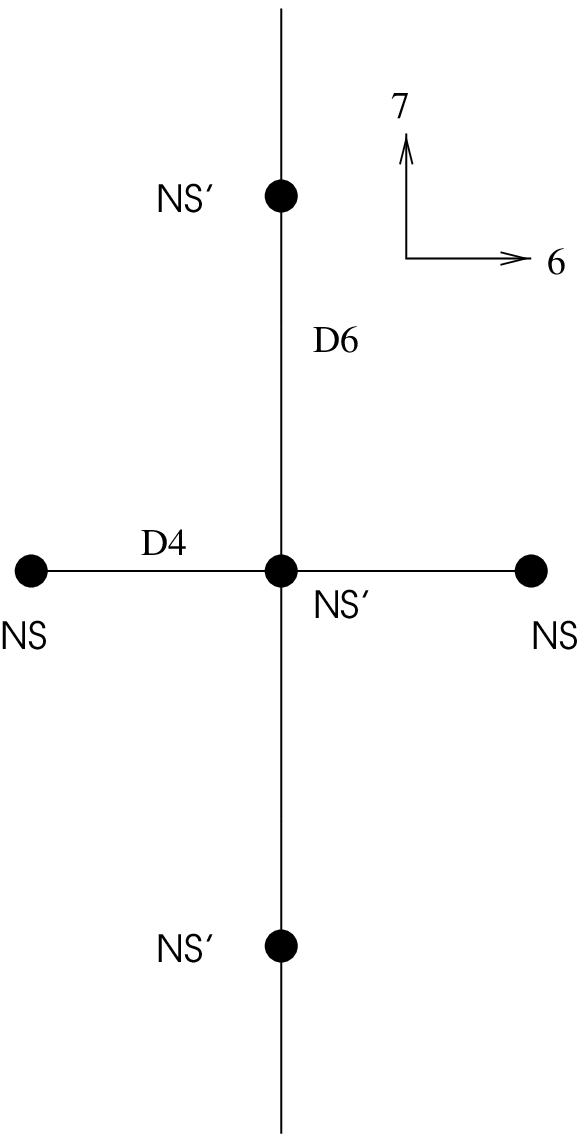}{4truecm}
\figlabel{\sixpuz}
\vspace{1cm}

\subsection{Flavor Doubling}
\label{cross}

Consider, as in figure \sixpuz, a system of NS-NS$'$-NS branes positioned in the
$x^6$ direction in
this order. Let $N_c$ D4 branes be stretched between the left NS brane and the
NS$'$ brane, and $N_c'$ D4 branes be stretched in between the right NS brane and
the NS$'$ brane. Let $N_f$ D6 branes be positioned at the $x^6$ position of the
NS$'$ brane, in such a way that they touch the D4 branes and give rise to
massless multiplets. This is precisely the ``cross" configuration, figure
\cross, studied in the
previous section. Using the results from this section, the gauge theory and
matter are
$SU(N_c)\times SU(N_c')$ with matter $4N_f$ chiral fields:
$N_f$ flavors $\tilde Q, Q$ in $(1,N_c')$, and $N_f$ flavors $\tilde R, R$
in $(N_c,1)$. There are two bi-fundamental chiral fields $\tilde F, F$ in the
$(N_c,\overline{N_c'})$ and its complex conjugate, $(\overline{N_c},N_c')$.
There is a superpotential of the form
\beq
\tilde Q\tilde F R - \tilde R F Q.
\eeq
As usual, the 45 positions of the D6 branes give rise to masses for both
types of flavors. To do that, the global symmetry is broken explicitely to
$SU(N_f)$. The brane interpretation is clear. We reconnect the upper and lower
D6 branes to form a single D6 brane which moves in the 45 direction and gives
masses which transform in the adjoint representation of the new $SU(N_f)$ group.
A more confusing issue is with the $x^6$ motion of the D6 branes.
It is easy to see that such a motion to the left (right) gives mass to the
$R (Q)$ fields and leaves the $Q (R)$ fields massless. How can this be included
in the superpotential?

To solve this puzzle, it is convenient to recall the six dimensional system
which controls the global symmetries of our system
\cite{bk,bk2,hz,hztwo}. This is a six dimensional
$SU(N_f) \times SU(N_f)$ with a matter $f$ in the bi-fundamental representation.
It is convenient to assume the existence of two NS$'$ brane far along the $x^7$
direction in both sides of the NS$'$ at the origin of $x^7$, as in figure
\sixpuz.

Then, anomaly cancellation in six dimensions requires additional $N_f$
fundamental fields for each $SU(N_f)$ group, as can be read off from the brane
configuration.

The main point in introducing the six dimensional theory is that its moduli
space is mapped to the space of parameters of the four dimensional theory.
Let us look at a particular configuration, in which an infinite D6 brane
reconnects and moves in the 456 direction. This corresponds to a flat direction
for the bi-fundamental field $f$.\footnote{The gauge invariant operator is
actually a meson which includes the lower and upper flavors of the six
dimensional theory, with the bi-fundamental fields.}
There are four scalars associated with such a motion.
These are the 456 distance to the NS$'$ brane and the reduction of the $A_7$
component of the gauge field which lives on the worldvolume of the D6 brane.
They decompose into
45 chiral field, $m$, and $\tilde m=x^6+A_7$ chiral field. As said above, the 45
expectation values are mapped to bare masses of both the $Q$ fields and the $R$
fields.
The puzzle raised in the last paragraph is now solved by introducing the
new mass term coming from the six dimensional $\tilde m$ chiral field.
D6 branes with positive $x^6$ position contribute to the superpotential
$\tilde m \tilde RR$. D6 branes with negative $x^6$ position contribute
$\tilde m \tilde QQ$ to the superpotential. We see that there are two patches
in which the superpotential looks different. The transition between the two
regions is smooth since the difference in the contribution to the
two superpotentials vanishes there.

We can generalize further the system by introducing $k$ left NS branes and $k'$
right NS branes. As suggested in \cite{EGK},
this includes the addition of two adjoint fields, $x$ and $y$,
for the left and right group, respectively.
The superpotential then becomes
\beq
x^{k+1}+y^{k'+1}+(x-y)\tilde FF+\tilde R(m-x)R+\tilde Q(m-y)Q+
\tilde Q\tilde F R - \tilde R F Q +W_p,
\eeq
where $W_p$ is defined in two different regions,
\begin{eqnarray}
W_p&=&\tilde m\tilde QQ,\qquad x^6<0 \\
&=& \tilde m\tilde RR,\qquad x^6>0
\end{eqnarray}
We use this superpotential in the next sections to determine more complicated
setups.

\subsection{Duality for the Product Gauge Group}
\label{product}

We proceed by recalling some of the properties in the models which were studied
in \cite{brodie}. The brane configuration is slightly more general than the one
introduced in the previous section. However, there is also a simplification.
For the moment, we avoid the issue of ``flavor doubling" and keep the D6 branes
away from the NS$'$ brane in the $x^6$ position.
The D6 branes give rise to flavors of one of the
gauge groups, not both, depending on their $x^6$ position.

\vspace{1cm}
\fig{
Notation for the angles in the product group setup.}
{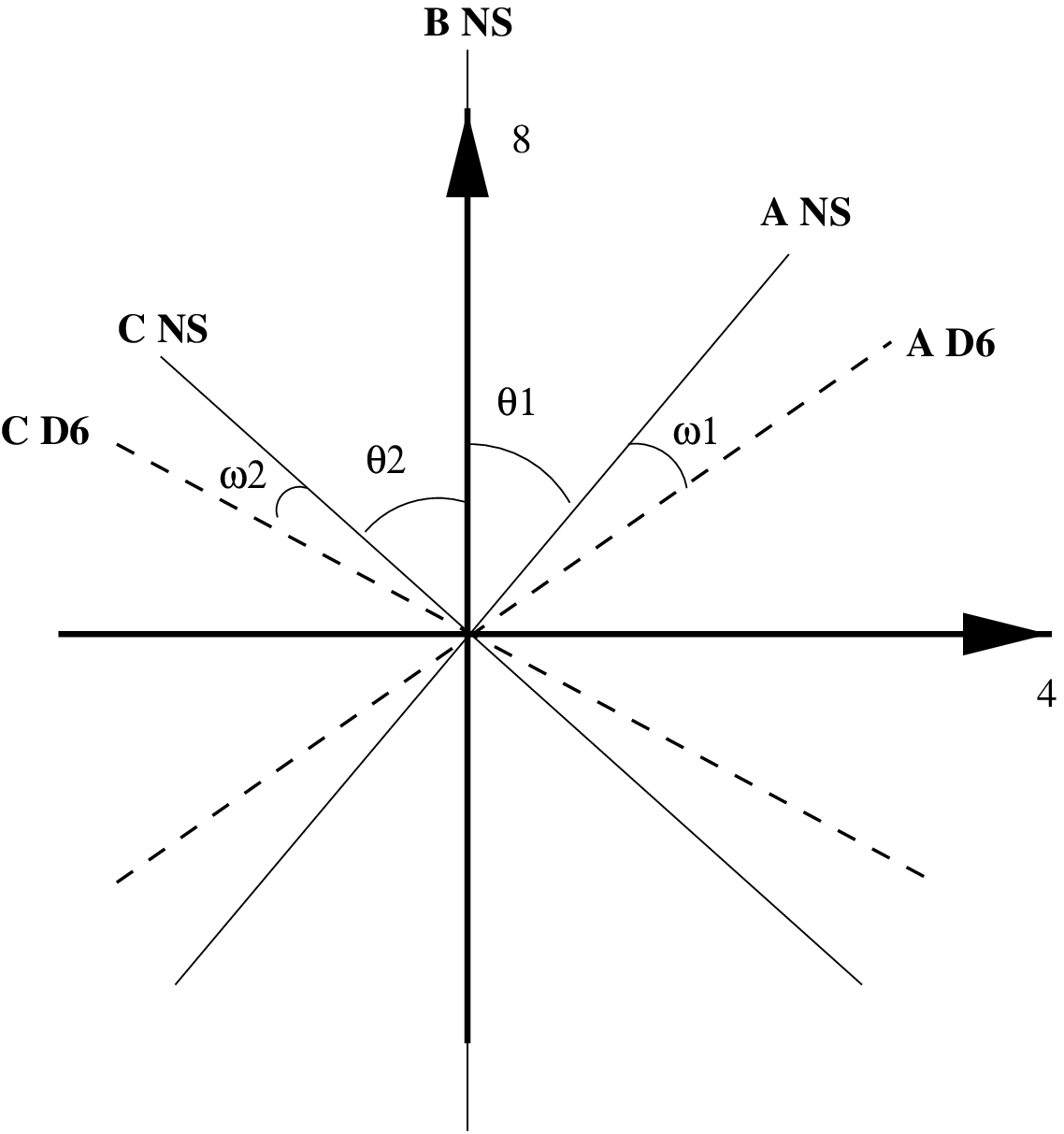}{7truecm}
\figlabel{\notation}
\vspace{1cm}

First, we allow for a general number of NS, NS$'$ and NS branes. For
convenience, we will denote them A, B and C, respectively.
Second, we will allow for general angles in the 45-89 space. The names NS brane
and NS$'$ brane are no longer useful.
Instead we use the general name NS branes, keeping in mind that each NS brane
has a general angle which specifies the orientation in 45-89 space.
The configuration is depicted in figure \notation.
In this figure, we also draw the angles for the various branes.
The B branes remain NS$'$ branes by a choice of the origin for the angular
variable.
Let $(k,k',k'')$ denote the number of A, B and C type NS branes respectively.
According to \cite{brodie} the $(k,k,k)$ case leads to a superpotential
\begin{equation}
\label{SUPO}
W=m_1 X_1^{k+1} + m_2 X_2^{k+1} + X_1 \tilde{F} F + X_2 \tilde{F} F +
\lambda_1 Q X_1 \tilde{Q} + \lambda_2 Q' X_2 \tilde{Q}' 
\end{equation}
while the $(k,1,k)$ case leads to
\begin{equation}
\label{SUPO2}
W=-\frac{1}{2} \left ( \frac{1}{m_1} + \frac{1} {m_2} \right ) 
( F \tilde{F})^{k+1}+
\lambda_1 Q X_1 \tilde{Q} + \lambda_2 Q' X_2 \tilde{Q}'.
\end{equation}
In the following, we identify these superpotentials by their triple number,
$(\cdot,\cdot,\cdot)$.
The coefficients in the superpotential are determined in terms of
the angles as
\begin{eqnarray}
\lambda_1 &=& \sin \omega_1 \\
\lambda_2 &=& \sin \omega_2 \\
m_1 &=& \tan \theta_1 \\
m_2 &=& \tan \theta_2.
\end{eqnarray}

To discuss the duality in these gauge theories it will be easiest to focus
on the case where $\omega_1=\omega_2=0$, as it was
discussed in \cite{brodie}. Choosing non-zero values for these angles
amounts to turning on the $Q X_1 \tilde{Q}$ perturbation.

The magnetic theory is obtained from this by switching the two end
NS branes while keeping the center NS brane in place. The resulting
gauge group is determined by charge conservation and can be read of
from the brane configurations using the linking numbers of \cite{HW}.

\vspace{1cm}
\fig{The product gauge group considered in \cite{brodie}. The thick lines are
NS branes at arbitary angle in 45-89 direction. The D6 branes are parallel to
the A and C NS branes.}
{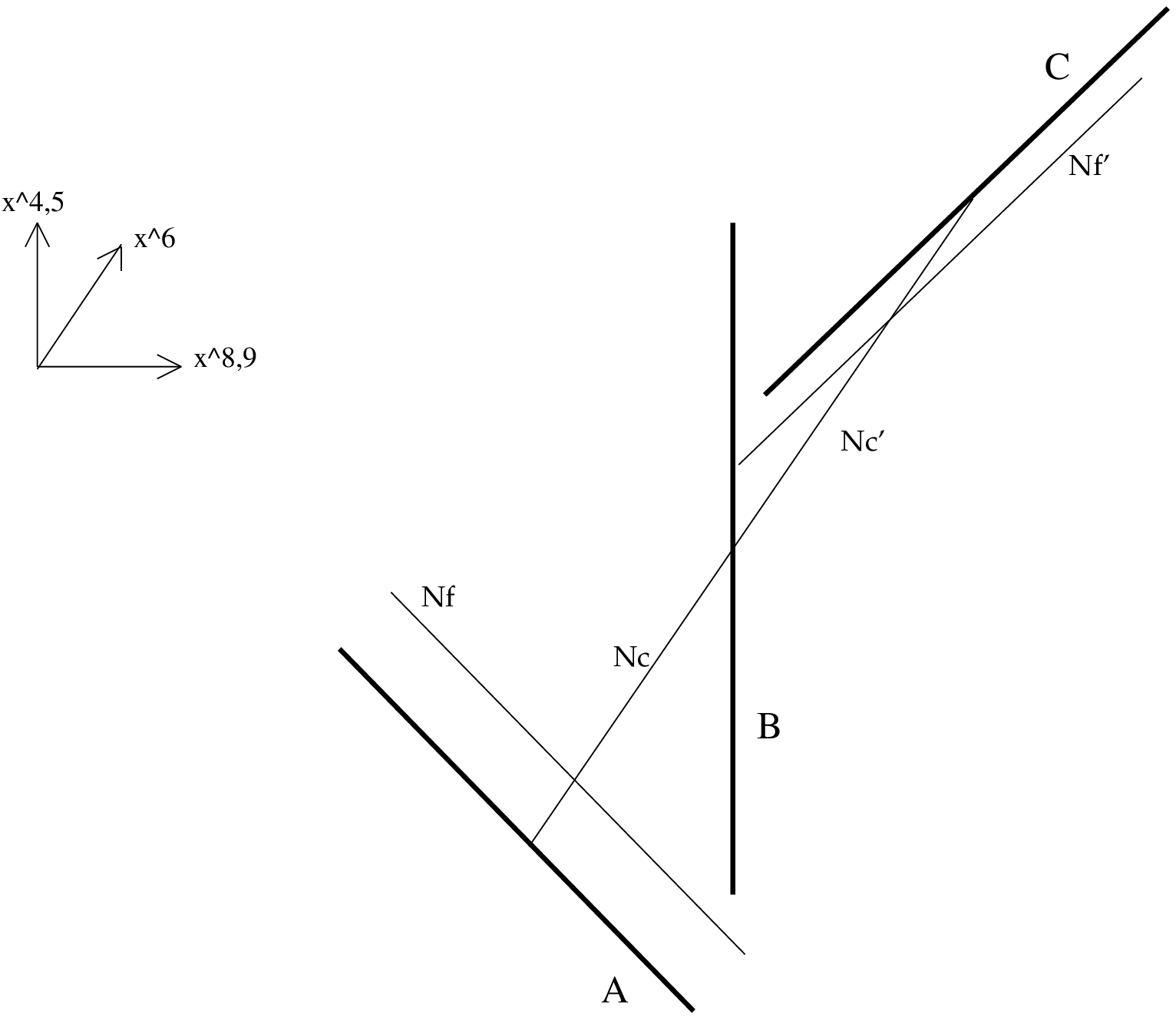}{7truecm}
\figlabel{\ils}
\vspace{1cm}

\vspace{1cm}
\fig{The dual of figure \ils, done by reversing the order of NS and D6 branes
\cite{brodie}.}
{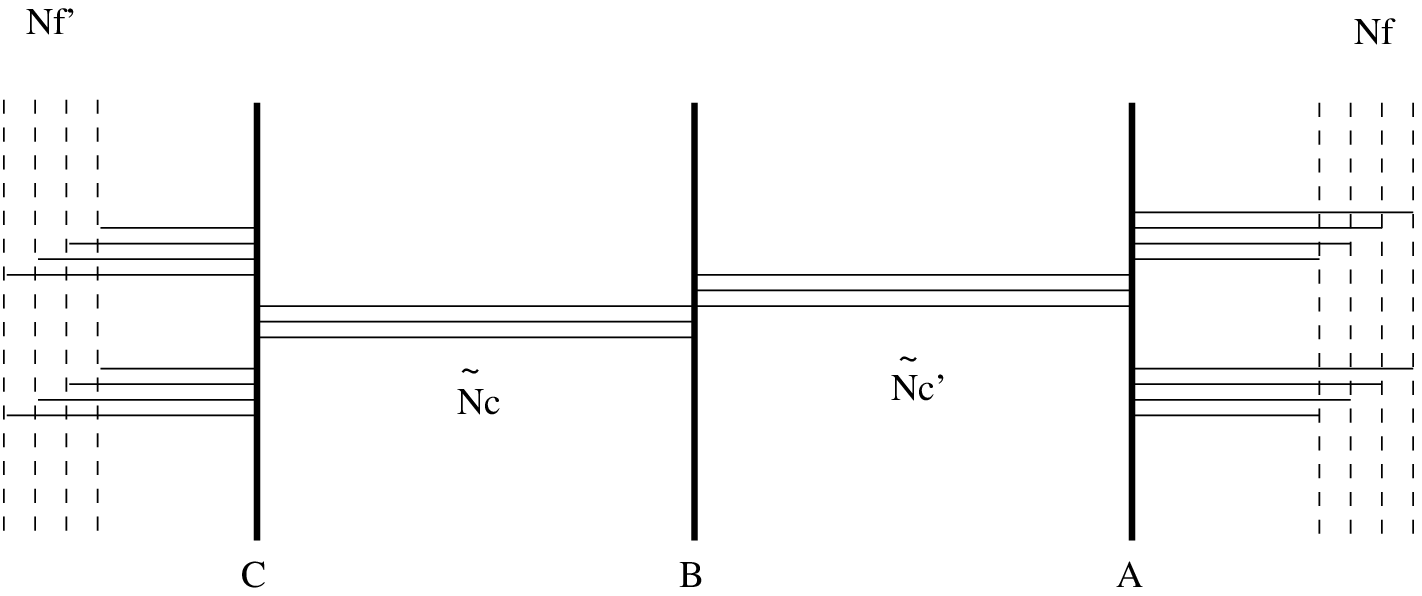}{7truecm}
\figlabel{\dual}
\vspace{1cm}

The linking number of a NS brane is given by $L_5 = \frac{1}{2}
(n_{6L}-n_{6R}) +(n_{4R}-n_{4L})$ and that of a D6 brane
by  $L_6 = \frac{1}{2}
(n_{5L}-n_{5R}) +(n_{4R}-n_{4L})$ where $n_{pL,R}$ denotes the number
of $p$ branes to the left(right) of the brane under consideration. However
parallel branes do not contribute to the linking number \cite{brodie}.
With this, it is easy to find that for $(k,1,k)$ the
dual gauge group is $SU(k N_f + (k+1) N'_f - N'_c) \times 
SU(k N'_f + (k+1) N_f - N_c)$ and for (k,k,k) the dual gauge
group is $SU(k N_f + 2k N'_f - N'_c) \times
SU(k N'_f + 2k N_f - N_c)$. For general $(k,k',k'')$ the dual
is $SU(k'' N_f + (k+k') N'_f - N'_c) \times
SU(k N'_f + (k+k'') N_f - N_c)$, even though the field theory interpretation
of this brane configuration is not clear at this stage.

\section{Chiral Gauge Theories}

We now turn to study a particular chiral gauge theory. All the information
collected so far in the previous sections will prove to be useful in studying
this system. Let us first give the general setup.

\subsection{The Setup}

Let us consider the same setup as in the previous section. In addition 
we include an orientifold 6 plane. As with the D6 and NS branes supersymmetry
allows for O6 planes with arbitrary orientation in the 4589 direction. Using
the by now familar notation we will call an orientifold parallel to the D6$'$
brane (that is stretching along 0123457) an O6$'$ and an orientifold
parallel to the D6 an O6. To fix the overall orientation we will always take
the center (B) brane to be an NS$'$ brane. In order to be able to introduce
the orientifold plane, the setup has to be symmetric with respect to
the orientifold. This means in particular:

\begin{itemize}
\item $N_c=N'_c$
\item $N_f=N'_f$
\item $\theta_1=-\theta_2$ and $\omega_1 = -\omega_2$ 
(the A branes have to be mirrors of the C branes)
\item the number of A branes and C branes has to be equal,
that is we only consider
$(k,k',k)$ configurations
\item one can only include an O6 or an O6$'$ (the B NS$'$ brane, carrying one
unit of NS charge,  has to be
self-mirror)
\end{itemize}

A different way to project to $SO$ and $Sp$ groups would be to introduce
an O4 plane. This possibility was discussed in \cite{ahn}.

\subsection{Identification of the gauge group}

Consider the following setups involving O6 planes. 
\vspace{1cm}
\fig{Theories with orientifolds and their deformations}
{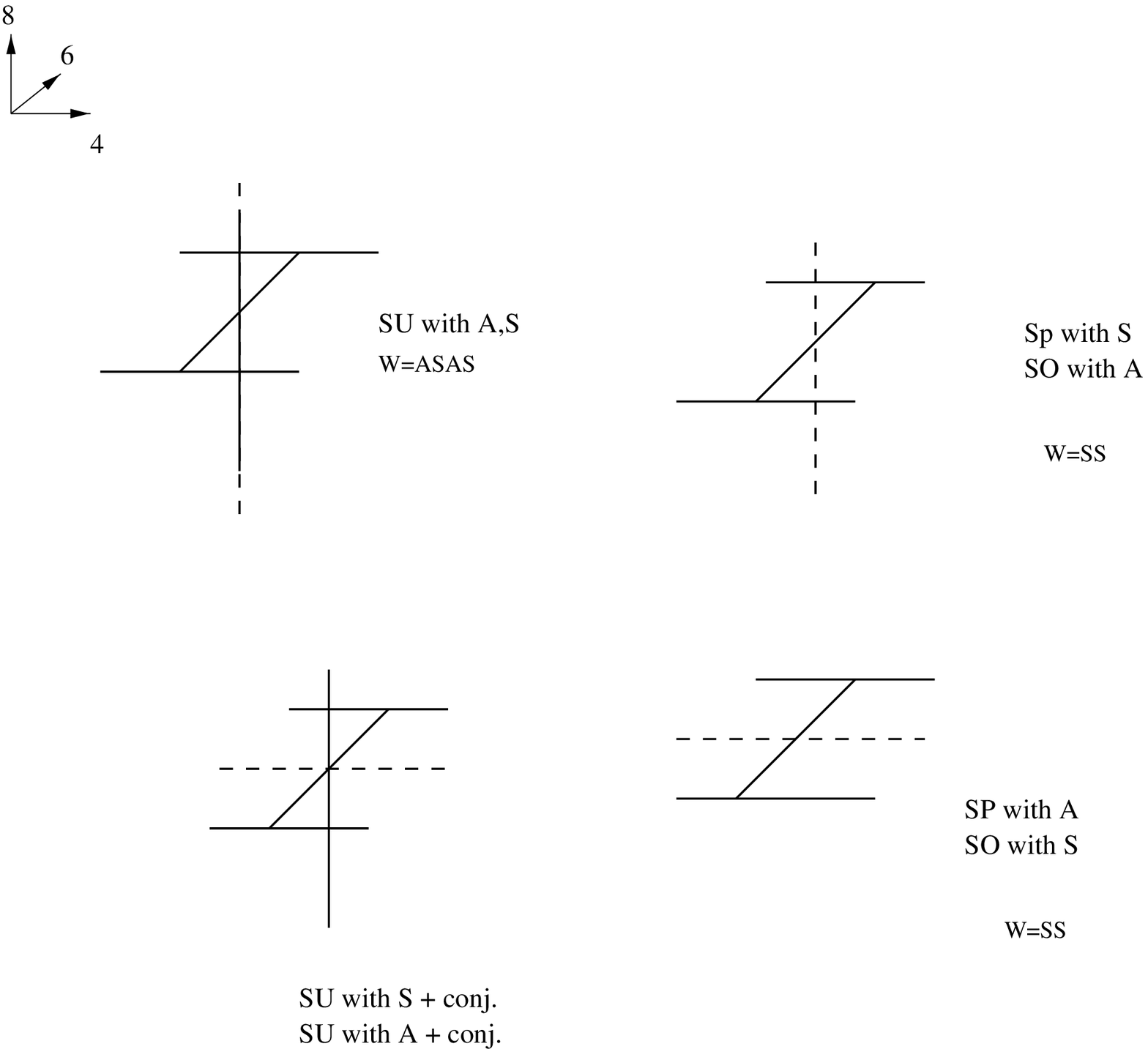}{14truecm}
\figlabel{\seat}
\vspace{1cm}
On the left in figure \seat we find the two possible projetions
of the product gauge group setup, the right hand side displays
two possible setups involving just 2 NS branes and an orientifold, which
arise as possible deformations of the theories on the left, as we will see
in the following. Let us briefly summarize the resulting gauge groups
and matter contents. A more detailed discussion of this identification
will be presented in the following sections. 

In the two pictures on the left the two $SU$ factors are identified
under the orientifold projection. One adjoint field is present, whose
mass is given by the angle $\theta$ between NS and NS$'$ brane. In
addition there are degrees of freedom that gave rise to
bifundamentals in the product gauge groups. In the following
two sections we will show that in the orientifolded theories
they will give rise to a complex symmetric and an antisymmetric tensor
for the O6 and one flavor of symmetric or antisymmetric tensors for the
O6$'$, depending on the sign of the projection.

The configuration in the upper right corner of figure \seat
is just the $N=2$ setup analyzed in \cite{LL}. The corresponding
gauge group is $SO$ ($Sp$) depending on the sign of the
orientifold projection. Rotating the NS branes breaks
$N=2$ to $N=1$ by giving a mass to the adjoint chiral multiplet
in the $N=2$ vector multiplet. We are left with an $SO$ ($Sp$).
Note that this mass is
already infinite at $\theta=\frac{\pi}{4}$, 
since this time it is given by the angle 
between
the outer NS branes which is twice the angle $\theta$ 
between outer NS and NS$'$ that
determined the mass before. Rotating further we claim that instead of
the adjoint tensor that became infinitely heavy a new tensor
with the opposite symmetry properties is coming down from infinite mass.
We will find evidence for this from comparing the branches of
the various brane configurations with results from field theory.
Note that if we were dealing with D branes instead of the NS branes
the rotation we performed would precisely change their worldvolume
gauge theory from $SO$ to $Sp$. It's therefore reasonable to
assume that something similar happens to the NS branes, too.
This way we identify the gauge group corresponding to the brane
configuration in the lower right corner of figure \seat to
be a $SO$ ($Sp$) gauge theory with a symmetric (antisymmetric) tensor.
Note that we this way reproduced all the $A_1$ theories in
the classification of \cite{brodiestrass}.

\subsection{$SU$ gauge groups with 2-index tensors and adjoint}

\subsubsection{The brane configuration}

Let us first discuss the case of an O6$'$ plane. We have to distinguish
two cases, since the orientifold can carry two different charges, leading
to different projections. The orientifold projection leads to the following
gauge group and matter content: since the left gauge group is identified
with the right gauge group, we are only left with one $SU(N_c)$
gauge group. The two adjoint fields are also identified and we are only
left with one adjoint $X=X_1=X_2$. 
Some of the degrees of freedom that  lead to the bifundamentals 
are projected out since the 
corresponding strings stretch across the orientifold, leaving
a single flavor of a symmetric or an antisymmetric tensor (that is
the tensor and its complex conjugate).
The superpotential for an O6$'$ for $(k,k,k)$ with
a positive RR charge can
easily be obtained from (\ref{SUPO}) to be
\begin{equation}
W= m X^{k+1} + \lambda Q X \tilde{Q} + X S \tilde{S}
\end{equation}
where $m=m_1$ and $\lambda=\lambda_1$ can be obtained from
the angles as above and $S$, $\tilde{S}$ denotes the symmetric
flavor obtained from the bifundamentals after projection.
The same superpotential with $A$ substituted for $S$ is obtained
if we use the other projection, with negative O6$'$ charge,
to obtain the antisymmetric tensor.

\subsubsection{Flat Directions}

We will only discuss the case with the symmetric flavor (corresponding to the
O6$'$ with positive charge), the antisymmetric
tensor can be treated in a completely analogous fashion, replacing $SO$
with $Sp$ groups. The field theory was analyzed in \cite{intri} for the
superpotential (1,1,1) and in \cite{brodiestrass} for the superpotential
$(k,k,k)$. For simplicity we will
focus just on the (1,1,1) case. We will only consider the configuration
with $\lambda=0$.
First consider the case with $\theta=\pi/2$. The adjoint becomes infinitely
heavy and hence all quartic superpotentials vanish.
In the brane picture the A and C branes are parallel. This time,
the D4 branes are free to move in the 4,5 direction along the A and C
NS branes. In the field theory this new branch opens up since
the $(S \tilde{S})^2$ term vanishes and now $S$ and $\tilde{S}$ are free to get
vevs of the form
$$\left<S\right> =  \left (
           \begin{array}{cccc}
                         x_1 & & & \\
                                & x_2 & & \\
                                       & & \ddots & \\
                                            & & & x_n
          \end{array} \right),
$$
and
$$\left<\tilde S\right> =  \left (
           \begin{array}{cccc}
                         \tilde{x_1} & & & \\
                                & \tilde{x_2} & & \\
                                       & & \ddots & \\
                                            & & & \tilde{x_n}
          \end{array} \right)
$$

\vspace{1cm}
\fig{Patterns of flat directions for the (1,1,1) theory with no superpotential.
In figure A we see the motion of D4 branes along the NS branes in the 45
directions which leads to either $SU$ or $SO$ theories. In figure $B$ we see
a slice in the 7 direction in which the NS branes may be moved away from the
NS$'$ branes, leading to a non-zero constant, $r$.}
{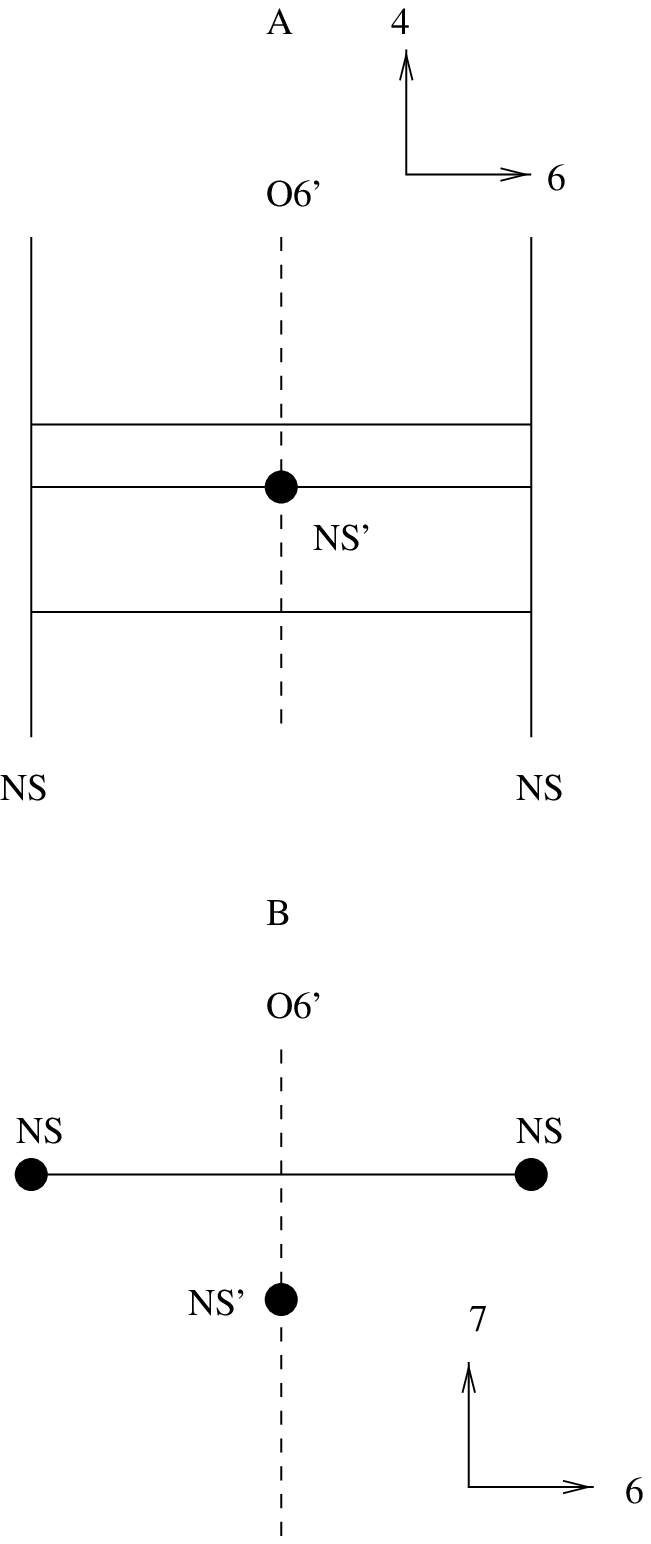}{6truecm}
\figlabel{\sym457}
\vspace{1cm}

with $|x_i|^2 - | \tilde{x}_i|^2 = r$, $r$ being a constant,
generically breaking $SU(N_c)$ completely \cite{intri}.
The $z_i=x_i \tilde{x}_i$ label the 4,5 position of the $i$-th brane.
If $n$ branes have $z_i=0$ they coincide at the origin (the NS$'$
position) and yield an unbroken $SU(n)$ gauge group with the original
matter content. If $p$ coincide at non-zero $z$, they give an unbroken $SO$
subgroup, which is obvious from both brane and field theory points of view.
See figure \sym457 A. The constant $r$ is identified with the distance between
the NS branes and the NS$'$ brane, as can be seen in figure \sym457 B.

For general values of $\theta$ the mass term for $X$ and the resulting
quartic superpotential, $(\tilde SS)^2$,
do not allow for any Coulomb branch. However there
are several Higgs branches. The mesonic branches arise in the standard
fashion by splitting the D4 branes between the D6 branes. The gauge
group is broken to the same model with a lower rank gauge group.

More interesting are the baryonic branches of the theory. In the
brane picture baryonic branches are associated with motion of NS
branes in the 7 direction. 
Consider the flat direction along which the baryon
$S^n Q^{N_c-n} Q^{N_c-n}$ gets
an expectation value. The field theory analysis was performed
in \cite{intri}. Along this flat direction the gauge group is
broken to $SO(n)$ with a symmetric tensor and
a superpotential $W=m X^2$. What happened
in the brane language is that we moved the center NS$'$ brane along the 7
direction. Thereby we obtain a configuration of parallel NS branes with
a parallel orientifold in between, giving rise to an $SO$ gauge
group with a symmetric tensor as claimed. 

\subsubsection{Non-Abelian Duality}

Similarly, one can carry over the construction of the non-Abelian duality from 
the discussion of the product gauge group in section \ref{product}.
Consider the (1,1,1) case. 
As above, the dual gauge group
is most easily obtained for the case $\lambda=0$, that is with the D6 branes
parallel to the outer NS branes, $\omega=0$. 
The only difference is that we have to take into account the orientifold plane.
Since it is charged like $\pm 4$ D6 branes we get an additional $\pm 4$
in the dual gauge group. Since we can not move the O6 to the outside
of the NS branes like we did with the D6 branes, the O6 charge appears
without the factor of 3 and we get as the dual gauge group an
$SU(3 N_f \pm 4 - N_c)$ gauge group with the same charged matter content
and several meson fields in agreement with field theory expectations. 

If we move the D6 branes on top of the orientifold and hence also on top
of the center NS brane, the issue of flavor doubling as before arises.
For a positive charge O6$'$, with the additional terms in equation (\ref{supo}),
the superpotential takes the form
\begin{equation}
W= S \tilde{Q} \tilde{Q} + \tilde{S} Q Q
\end{equation}
where we now have $2 N_f$ $Q$ fields instead of the $N_f$ $Q$ fields
considered so far. The global symmetry is enhanced from $SU(N_f) \times
SU(N_f)$ to $SO(2 N_f) \times SO(2 N_f)$ as expected for D branes coinciding
on top of an orientifold.
Similarly, for a negative charge O6$'$, $S$ is replaced by $A$ and the global
symmetry is enhanced to $Sp(N_f)\times Sp(N_f)$.

\subsection{The Chiral Theory}
\label{chiral}
\subsubsection{The brane configuration}
\vspace{1cm}
\fig{An O6 plane and a NS$'$ brane, the ``fork".
The sign of the O6 is changed as it crosses
the NS$'$ brane. Dashed line represents O6 with charge +4. Dotted line
represents O6 with charge -4.
Anomaly cancellation requires the existence of semi-infinite D6 branes.
The 4 D6 branes and their images, are represented by fat solid lines.}
{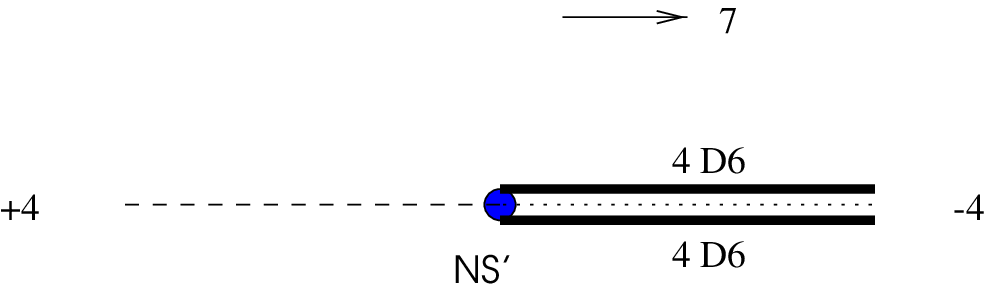}{10truecm}
\figlabel{\beast}
\vspace{1cm}

The story becomes more complicated once we consider the case of an O6 plane.
As considered in \cite{johnson} an O6 changes sign when passing through an
NS$'$ brane.
Anomaly freedom of the corresponding 6d field theory requires this object
to be accompanied with 4 physical half D6 branes and their mirrors stuck
on the negatively charged half of the O6. In total we have the following
picture, as in figure \beast,
(reading along the 7 direction): from infinity running inwards
we have an O6 of charge +4. At $x_7=0$ we find the NS$'$ brane with the system
of D4 branes attached to it. By passing through the NS$'$ the O6 changes
sign. In addition 4 half D6 branes and their mirrors start out from
here and stretch to infinity. They also have a total D6 charge of +4. It
therfore seems to be reasonable to assume, that this whole system has 
a smooth M-theory interpretation. However this interpretation is still
to be found. The half D6 branes are stuck at the orientifold.

As above, this system can be most easily analyzed by considering it as
a projection of the product gauge group setup. Again the orientifold
identifies the two gauge groups and the two adjoints. The degrees of
freedom that lead to the bifundamentals
turn into one antisymmetric and one conjugate symmetric tensor,
since this time we have a different projection acting on the different sides
of the D4 branes. In addition we have matter coming from the half D6 branes.
Since they are stuck at the orientifold and the orientifold is stuck on top
of the center NS$'$ brane, we necessarily have the half D6 branes on top
of the NS$'$ brane and hence the issue of flavor doubling arises. 
This way we get an $SU(N_c)$ gauge group with one antisymmetric
tensor $A$,
one conjugate symmetric tensor $\tilde{S}$ and 8 fundamentals $T$. This gives
rise to an anomaly free chiral theory! In addition there
are $N_f$ vectorlike fundamental flavors $Q$, $\tilde{Q}$ 
arising from the D6 branes.

For simplicity we will only consider the $(1,1,1)$ case, that is
only one NS brane of each type.
The superpotential is 
\begin{equation}
W=\tilde{S} T T + \lambda X Q \tilde{Q} + m X^{2} + X A \tilde{S}
\end{equation}
where the first term is the superpotential required by the doubling
mechanism, equation (\ref{supo}) and the rest is the remainig superpotential
from the product gauge group (\ref{SUPO}) after projection.
$m$ and $\lambda$ are again given in terms of the angles as above.
For nonzero $m$ we can integrate out the adjoint and obtain
quartic superpotential terms instead.

If we choose to move the additional D6 branes on top of the orientifold
we again have to deal with doubling. From the brane configuration
it is obvious that this way the global symmetry is enhanced to $SO(2 N_f +8)
\times Sp(N_f)$ if we are dealing with D6 branes and $SO(8) \times
SO(2 N_f) \times Sp(N_f)$ if we are dealing with D6$'$ branes. For
more general orientations of the D6 branes we are left with
an $SO(8) \times SU(N_f) \times SU(N_f)$ global symmetry. To achieve
this enhanced global symmetry in the field theory we add superpotential
terms corresponding to (\ref{supo}) with the coefficients depending on
the various rotation angles.

\subsubsection{Flat Directions}

The field theory considerations can again be found in \cite{intri,
brodiestrass}.
Since we focus on the $(1,1,1)$ case we are dealing with a superpotential
$$W= \tilde{S} T T + \lambda X Q \tilde{Q} + X A \tilde{S} + m X^2$$
with $m=\tan \theta$ and $\lambda = \sin \omega$ with the angles defined
as in figure \notation. Consider the case $\lambda=0$.
For non-zero $m$ we can integrate out the massive adjoint and
obtain
$$W= \tilde{S} T T - \frac{1}{2m} (A \tilde{S} )^2 $$
For $\theta=0, \pi/2$ the theory has a Coulomb branch. This is obvious
from the brane picture. For $\theta=0$ all branes are parallel. The D4
branes are free to move in the 8,9 direction. In the field theory for
$\theta=0$ the adjoint remains massless. It's vev parametrizes the
Coulomb branch.

For $\theta= \pi/2$ the A and C brane are antiparallel. This times
the D4 branes are free to move in the 4,5 direction along the A and C
NS branes. In the field theory this new branch opens up since
the $(A \tilde{S})^2$ term vanishes and now $A$ and $\tilde{S}$ are free to get
vevs of the form
$$\left<A\right> =  \sigma_2  \otimes \left (
           \begin{array}{cccc}
                         x_1 & & & \\
                                & x_2 & & \\
                                       & & \ddots & \\
                                            & & & x_n
          \end{array} \right).
$$
and
$$\left<\tilde S\right> = {\bf 1}  \otimes \left (
           \begin{array}{cccc}
                         \tilde{x_1} & & & \\
                                & \tilde{x_2} & & \\
                                       & & \ddots & \\
                                            & & & \tilde{x_n}
          \end{array} \right).
$$
with $|x_i|^2 - | \tilde{x}_i|^2 = r$,
where as above $r$ is associated with the position of the NS$'$ brane,
generically breaking $SU(N_c)$ to $U(1)^{\frac{N_c}{2}}$ \cite{intri}.
If $N_c$ is odd there is a zero in the lower right corner of $A$ and
$\tilde{S}$. Note that on this Coulomb branch as opposed to the one associated
with the adjoint, the rank of the gauge group is only half of the original
rank. In the brane picture this is obvious, since in the 4,5 plane the
orientifold is a point and hence the projection identifies a single brane
and its mirror. If $N_c$ is odd one of the D4 branes is stuck at the
orientifold.

For general values of $\theta$, the mass term for $X$ and the resulting
quartic superpotential don't allow for any Coulomb branch. However there
are several Higgs branches. The mesonic branches arise in the standard
fashion by splitting the D4 branes between the D6 branes. The gauge
group is broken to the same model with a lower rank gauge group.

More interesting are the baryonic branches of the theory. In the
brane picture baryonic branches are associated with motion of NS
branes in the 7 direction. Some of these baryonic branches already exist
for $N_f=0$ and describe the possible deformations of the ``fork"
(the sign changing orientifold with the embedded NS$'$ brane and the 8 half
D6 branes, figure \beast).
Consider the flat direction along which the baryon 
$A^n \tilde{Q}^{N_c-2n}$ gets
an expectation value. The field theory analysis was performed
in \cite{intri}. Along this flat direction the gauge group is
broken to $Sp(n)$ with a symmetric (adjoint) tensor and
a superpotential $W=m X^2$. What happened
in the brane language is that we moved the center NS$'$ brane along the 7
direction. Thereby we also moved the point at which the orientifold changes
sign. The NS - D4 system describing our remaining field theory now
only sees the negatively charged orientifold with the embedded D6 branes, 
leading to an
$N=2$ $Sp(n)$ gauge theory for the case that the A and C branes are both
NS. Rotating the A and C NS branes leads to the usual breaking
of $N=2$ to $N=1$ by introducing the mass for the adjoint scalar
in the $N=2$ vector multiplet.

Similarly we can move the NS$'$ brane in the other direction, leaving us
with an $SO(n)$ ($n<N_c$) gauge theory with an adjoint and
a superpotential $W=m X^2$ for this adjoint. In the field theory
this is also a baryonic flat direction, this time associated
with the baryon operator $\tilde{S}^n Q^{N_c -n } Q^{N_c -n}$.  
The 8 $T$ fields no longer contribute any massless matter to the brane
configuration, since our gauge system now is at a finite distance
from the half D6 branes. In the field theory this effect
is realized by the $STT$ superpotential, giving the $T$ fields
a mass proportional to the vev of $S$.

\subsubsection{Non-Abelian duality} 

Since in the discussion of non-Abelian duality only the charge of the
orientifold mattered, the brane picture seems to suggest that this
chiral theory is again described by a dual $SU(3 N_f +4 -N_c)$ gauge
group with the same charged matter content and the obligatory mesons, 
since the complicated object
stuck on top of the center NS$'$ brane has charge +4 on both sides, even
though this charge arises in a different fashion on both sides.
This kind of theory without the $\tilde{S} T T$ term
was also analyzed in \cite{brodiestrass}. There, the dual is found to be
an $SU(3 (N_f +4) - N_c)$ gauge theory. To compare with our field theory
prediction we have to analyze the effect of the $\tilde{S} T T $
perturbation in field theory. In the dual theory this operator
is mapped to a meson $M$ that couples to the charged magentic fields
via a superpotential \cite{brodiestrass}
$$
W= \ldots + M t \tilde{s} t + \ldots
$$
where the small letter fields transform under the same representations under
the dual gauge group as their capital counterparts under the original
gauge group.
Perturbing this superpotential with the operator $\tilde{S} T T$ means
that we add $M$ to the dual superpotential. Via the equations
of motion for $M$ this forces that $t \tilde{s}  t$ has non-
vanishing vev. This kind of vev was analyzed by \cite{pouliot}.
It was found that the dual group is broken by 1 unit
per $t$ flavor involved, that is in our case by 8 units.
The remaining dual gauge group is $SU(3 \, N_f +4 -
N_c)$ in perfect agreement with the brane prediction.

\subsection{Generalizations to Mutiple NS branes}

Similar to the product gauge group case we can generalize our setup
to $(k,k,k)$ or $(k,1,k)$. We would expect that this
corresponds to chiral theories with a superpotential (for $\omega=0$)
\begin{equation}
W=m X^{k+1} + X A \tilde{S} + \tilde{S} TT
\end{equation}
and
\begin{equation}
W=\frac{1}{2m} ( A \tilde{S})^{k+1} + \tilde{S} TT
\end{equation}
and similarly for the non-chiral theories discussed in section 4.3.
These theories correspond to the $D_{k+2}$ and $A_k$ models in
the classification of gauge theories with 2-index tensors of
\cite{brodiestrass}. The D-type model with $k=2$ and $N_f=N_c$
gives rise to a chiral finite theory and was analyzed in \cite{luest}.
A better understanding may shed some light on how finiteness
is visible in the $N=1$ brane setup.

Even though this approach seems very promising, there are some discrepencies
that need further understanding. To highlight this it is enough to consider
the easier case of $(k,0,k)$ as they arise on the baryonic branches
of our models. That is, for an O6$'$ we consider
a $SO(N_c)$ gauge group with an antisymmetric tensor, $N_f$ flavors
and superpotential $W=A^{k+1}$, while for the O6 we consider
a $SO(N_c)$ gauge group with a symmetric tensor, $N_f$ flavors
and superpotential $W=S^{k+1}$. These gauge theories arise on the baryonic
branch of the chiral and non-chiral models considered in section 4 respectively.
In the language of \cite{brodiestrass} they correspond to the
$A_k$ mezzanine and orchestra models respectively.
Similarly one could discuss the corresponding $Sp$ groups.

In the field theory analysis of \cite{intri,brodiestrass} it is
found that one should restrict oneself to $k$ odd in the case of $SO$
with the antisymetric tensor in order to truncate the chiral ring and get
a handle on the gauge theory. On the other hand for $SO$ with the symmetric
no such restriction exists. It is not clear how this translates into the brane
picture. This is a field theory prediction for the brane configuration which
deserves further study.

If one tries to describe non-Abelian duality the branes give us the prediction
that the dual gauge group should be $k N_f - N_c + const.$ where the
constant is associated to the orientifold. The field theory
tells us that the duals actually have $k N_f - N_c +4$ and $k N_f - N_c
+4k$ as their dual gauge theory. \footnote{Keep in mind that
for the mezzanine model
we relabelled 2(k+1) from the field theory to (k+1), since
this is the k that is associated with the number of 5 branes in the brane
picture}. Note that while we get the factor in front of $N_f$ right,
the orientation of the O6 affects the multiplicity
of the +4 in the dual gauge group.

The same comments apply to the other setups. The factors
of $(2k+1) N_f$ in the $A_k$ balcony (keeping the relabelling in mind) and 
and $SU$ mezzanine models of \cite{brodiestrass} as well
as the $3k N_f$ in the $D_{k+2}$ series can be easily obtained
from just translating the results of the product gauge group analysyis
of \cite{brodie}. 
However the effect of the orientifold charge and of the restrictions
on $k$ to be odd in some cases deserve a further investigation.

\section{Small Instanton Transition and Chiral non-Chiral Transition}

We now turn to a very interesting effect which arises in the brane
configuration. This effect was demonstrated in the context of six
dimensional gauge theories in \cite{hztwo} and was called a ``small instanton
transition". It demonstrates the effect in which a tensor multiplet is traded
with 29 hypermultiplets for a class of six dimensional models.
Following \cite{smallins,comments} 
this effect was first found in \cite{aspinwall} and further discussed in
\cite{ken}. We repeat the same brane motion for our four dimensional
system. The results are very striking. We get a transition between two models,
a chiral model and a non-chiral model.
Moreover, this result is generic, as any brane configuration for a chiral
theory has such a transition. We expect very interesting connections to arise
from this effect.

\vspace{1cm}
\fig{A small instanton transition which lead to chirality change in the
spectrum. A NS$'$ brane comes from infinity in the $x^7$ direction and combines
with the stuck NS$'$ at the origin. At this point they can both leave the
origin in the $x^6$ direction. The resulting four dimensional theory is no
longer chiral.}
{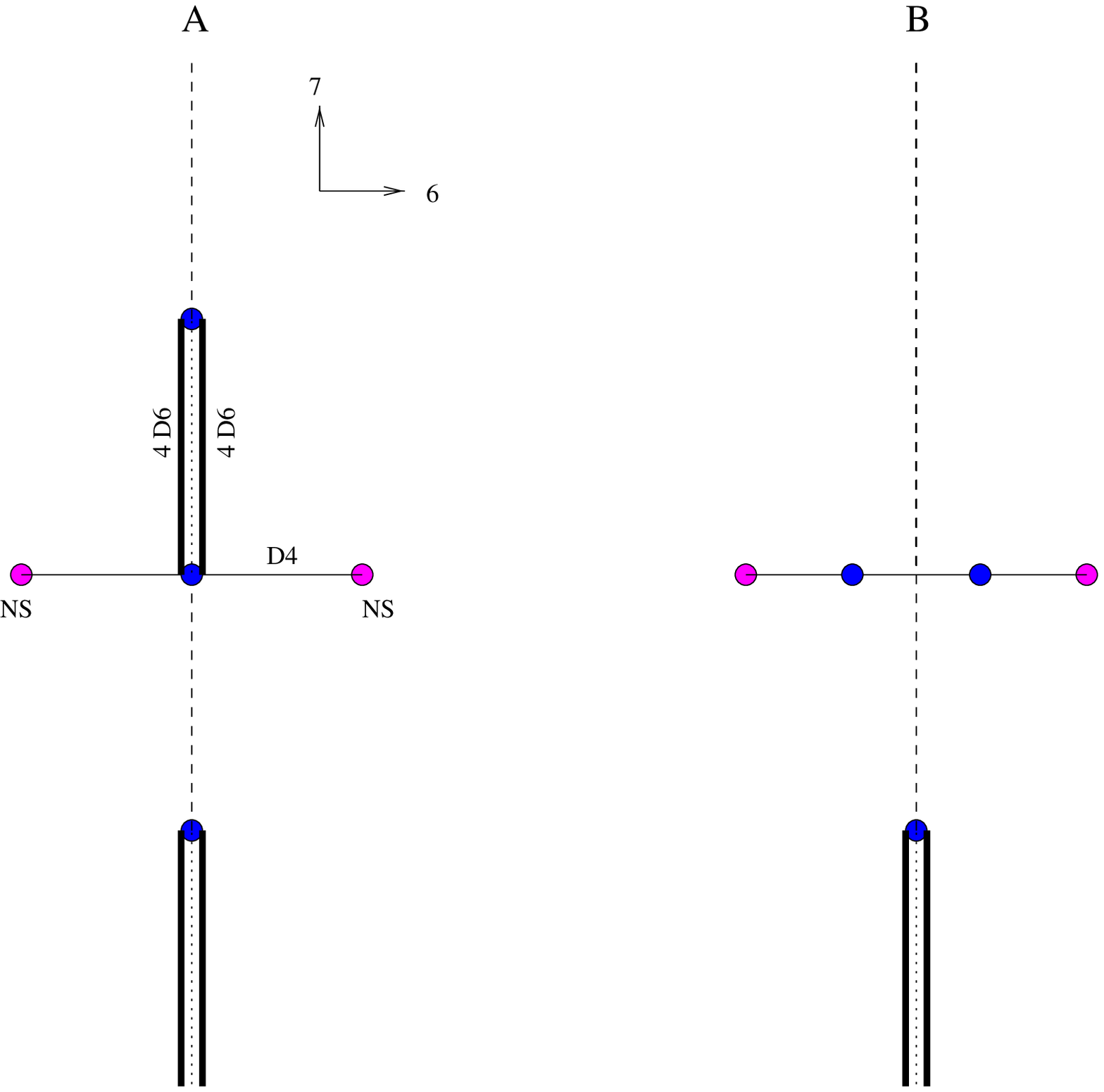}{10truecm}
\figlabel{\smallin}
\vspace{1cm}

Consider the configuration depicted in figure \smallin A.
It describes our chiral theory from section \ref{chiral}. Note, however that
there are two additional NS$'$ branes in the system. We will call them upper and
lower branes, respectively. The original NS$'$ brane, to which the four
dimensional system is attached, will be called central.
When these two NS$'$ branes are far away from the central NS$'$ brane, they can
not affect the dynamics of our theory and the model is just the model studied
in section \ref{chiral}.

As we did for the case with no orientifold plane, it is useful to think of the
underlying six dimensional theory which gives the global symmetries of the
four dimensional system. This theory can be read off from the brane
configuration and is just $SO(8)$. If, in addition, there are $N_f$ D6 branes
on top of the orientifold plane, the gauge group is $SO(8+N_f)\times Sp(N_f)$.
The matter content of this theory has one bi-fundamental hypermultiplet,
$N_f$ $(\fund,1)$ and $N_f+8$ $(1,\fund)$. Each gauge group couples to a tensor
multiplet which serves as a gauge coupling. The 
vev of the scalars in the two tensor multiplets determine
the distance between the two pairs of adjacent NS$'$ branes.

The NS$'$ branes are not allowed to move in the 456
directions. This is because they carry one unit of charge and the orientifold
allows for motion outside only in the case that an image exists.
 From a six dimensional point of view, this corresponds to the fact that there
are no FI terms for $SO$ or $Sp$ groups. On the other hand, the relative
motion of the NS$'$ branes along the $x^7$ direction corresponds to a change in
the two tensor multiplets in the thoery.
There is, however, an option for a pair of NS$'$ branes to move in the 456
directions. We can move two NS$'$ branes to touch in the $x^7$ direction.
At this point, the orientifold planes from above and below the pair of NS$'$
branes are identical, as is clear from figure \smallin.
 From a six dimensional point of view such a motion corresponds to taking one of
the gauge couplings to infinity and thereby obtain
 strings with vanishing tension - a non
trivial fixed point.
The pair of NS$'$ branes can then move in the 456 direction, as in figure
\smallin B.
The resulting six dimensional gauge group is now completely broken.

Let us go back to reinterpret this transition in our four dimensional system.
When the NS$'$ branes are far, we have our chiral theory.
When the NS$'$ branes move in the 456 directions, the theory is different.
We can read it off from figure \smallin B. We have $SO(N_c)\times SU(N_c)$ with
a symmetric tensor for the $SO$ group and a
pair (chiral and its conjugate) of bi-fundamental fields.
The symmetric tensor arises as described in section 4.2. In addition
we expect a superpotential $W=S^2 + S \tilde{F} F$.

This is done when we take the upper NS$'$ brane and move out with the central
NS$'$ brane. We can do it, instead, with the lower NS$'$ brane.
The resulting gauge theory
is $Sp({N_c\over2})\times SU(N_c)$ with an antisymmetric tensor 
for the $Sp$ and 
a bi-fundamental pair.
For this case, note that $N_c$ must be even, otherwise the motion of the NS$'$
branes to the 456 directions is not allowed and possibly breaks supersymmetry.

Here, we specify the matter content for the case when $N_f=0$. When $N_f\not=0$,
we need to take into account the effect of flavor doubling as discussed in
section \ref{cross}. It is straight forward and we will not repeat it here.
There is however a difference for the $Sp$ case. The upper 8 stuck D6 branes
are combined with the lower 8 stuck D6 branes. At this point they are no longer
stuck and give rise to 4 physical D6 branes, free to move in the 456 directions.
The number of flavors for the
$Sp({N_c\over2})\times SU(N_c)$ group is increased from $N_f$ to $N_f+4$.

To summarize, the analog of the six dimensional small instanton transition
gives rise to a transition between a chiral theory with a 
$SU(N_c)$ gauge theory with an antisymmetric tensor, 
one conjugate symmetric tensor and 8 fundamentals to either
\begin{itemize}
\item a $SO(N_c)\times SU(N_c)$ gauge group with matter transforming
as $(\fund \fund,1)$, $(\fund,\fund)$ and $(\fund, \overline{\fund})$
\item or a $Sp({N_c\over2})\times SU(N_c)$ gauge group with matter
transforming
as $(\asym,1)$, $(\fund,\fund)$, $(\fund, \overline{\fund})$ and 8 $(\fund,1)$.
\end{itemize}
In addition one can include an arbitrary number of flavors from D6 branes.
At the point in the transition when all the NS$'$ branes meet we find
in the field theory a non-trivial fixed point with enhanced global symmetry.
In the first case the global symmetry at the fixed point solely
comes from the coinciding NS$'$ branes. In the second case in addition to
this the $SO(8)$ flavor rotations are enhanced to $SO(8) \times SO(8)$ 
at the fixed point.

Acknowledgements

We would like to thank John Brodie and Alberto Zaffaroni for useful discussions.
A.H. would like to thank the Institut f\"ur Physik at the 
Humboldt Universit\"at
for their hospitality while this work was being completed.
The research of A.H. is supported in part by NSF grant PHY-9513835
and the research of I.B. and A.K. is supported by Deutsche
Forschungsgemeinschaft.

\bibliography{chiral}
\bibliographystyle{utphys}

\end{document}